\documentclass{iau} 
\usepackage{natbib}
\usepackage{graphicx}
\def\lya{Ly$\alpha$}

\def\apj{ApJ}
\def\apjl{ApJL}
\def\aap{A\&A}

\def\solphys{Sol. Phys.}

\title[EUV Irradiance Observations from SDO/EVE as a Diagnostic of Solar Flares] 
{EUV Irradiance Observations from SDO/EVE as a Diagnostic of Solar Flares}

\author[Ryan O. Milligan]   
{Ryan O. Milligan$^{1,2,3}$}

\affiliation{$^1$Astrophysics Research Centre, School of Mathematics \& Physics, Queen's University Belfast, University Road, Belfast, Northern Ireland, BT7 1NN \\email: {\tt r.milligan@qub.ac.uk}\\[\affilskip] $^2$Solar Physics Laboratory (Code 671), Heliophysics Science Division, NASA Goddard Space Flight Center, Greenbelt, MD 20771, USA\\[\affilskip] $^3$Department of Physics, Catholic University of America, 620 Michigan Avenue, Northeast, Washington, DC 20064, USA}

\pubyear{2015}
\volume{320}  
\jname{Solar and Stellar Flares and Their Effects on Planets}
\editors{Alexander Kosovichev, Suzanne Hawley \& Petr Heinzel, eds.}
\begin{document}

\maketitle

\begin{abstract}
For the past six years, the EUV Variability Experiment (EVE) onboard the Solar Dynamics Observatory has been monitoring changes in the Sun's extreme ultraviolet output over a range of timescales. Its primary function is to provide measurements of the solar spectral irradiance that is responsible for driving fluctuations in Earth's ionosphere and thermosphere. However, despite its modest spectral resolution and lack of spatial information, the EVE spectral range contains many lines and continua that have become invaluable for diagnosing the response of the lower solar atmosphere itself to an injection of energy, particularly during a flare's impulsive phase. In addition, high temperature emission lines can also be used to track changes in temperature and density of flaring plasma in the corona. The high precision of EVE observations are therefore crucial in helping us understand particle acceleration and energy transport mechanisms during solar flares, as well as the origins of the Sun's most geoeffective emission. 
\end{abstract}

\firstsection 
\section{Introduction}
Since its launch in February 2010, the Solar Dynamics Observatory (SDO; \citealt{pesn12}) has been providing us with an uninterrupted view of our Sun at extreme ultraviolet (EUV) wavelengths. This component of the Sun's output is a known driver of fluctuations in planetary atmospheres and so understanding its origins and variations - particularly during periods of extensive activity - remains a high priority for heliophysics research. The EUV Variability Experiment (EVE; \citealt{wood12}) onboard SDO was designed to measure changes in the Sun's irradiance at geoeffective wavelengths on timescales from seconds to years. The EUV emission detected by EVE (65--1050\AA), which affects the Earth's upper ionosphere (F-layer; $>$150 km) and thermosphere, contains many lines and continua that are generated in the chromosphere that become greatly enhanced during solar flares. The optically thick chromospheric Lyman-alpha line of hydrogen (hereafter \lya), on the other hand, is responsible for generating the ionospheric D-layer (80--100 km; \citealt{tobi00}), along with optically-thin soft X-ray (SXR) emission. 

Enhancements in chromospheric emission at flare footpoints and ribbons are believed to be driven by a `beam' of high-energy electrons accelerated from an energy release site in the corona, but details of how this energy is transferred to the chromosphere, and at what depth it is deposited, are still poorly understood. As well as being a primary driver of space weather, the EVE wavelength range contains a wealth of diagnostic tools for probing the plasma conditions during solar flares themselves. Such diagnostics include temperature and density sensitive line ratios \citep{mill12b}, plasma flow velocities \citep{huds11}, differential emission measures and emission measure distributions \citep{kenn13}, energetics \citep{mill14}, continuum enhancements and temperatures \citep{mill12a,mill14}, elemental abundances \citep{warr14}, heating and cooling rates \citep{mill15b,cham12}, and continuum contributions to broadband imagers \citep{mill13}. For further reading on EUV spectroscopy of the flaring solar atmosphere see \cite{mill15a}. In this paper I shall highlight some of the insights into solar flares that have been gained using EVE data beyond its primary science goals. 

\section{The EUV Variability Experiment}
\label{sec:eve}

SDO/EVE acquires full-disk, Sun-as-a-star EUV spectra every 10 seconds making it ideal for studying the temporal evolution of solar flare emission. The MEGS-A (Multiple EUV Grating Spectrographs) component covered the 65--370\AA\ wavelength range, which includes the He~II 304\AA\ resonance line and He~II free-bound continuum with a recombination edge at 228\AA, with a near 100\% duty cycle. Unfortunately, MEGS-A is currently no longer taking data due to a power anomaly that occurred on 26 May 2014. The MEGS-B component (370--1050\AA) observes the H~I continuum (Lyman; hereafter LyC) and He~I continuum, with recombination edges at 912\AA\ and 504\AA, respectively, as well as numerous emission lines formed over a broad range of temperatures, while the MEGS-P broadband ($\sim$100\AA) diode is centered on the \lya\ line of hydrogen at 1216\AA. Despite the loss of MEGS-A, MEGS-B and -P continue to operate, albeit at a reduced duty cycle due to unforeseen instrumental degradation. In October 2015, the EVE flight software was updated so that MEGS-B and -P will now respond autonomously to M-class flares or greater based on the SXR flux detected by the ESP 1--7\AA\ channel. Multiple 3-hour flare campaigns are permitted each day, and a 3-hour observation is made each day even if a flare campaign had not been triggered. The decision to prioritize exposing MEGS-B and -P during periods of increased activity during the extended phase of the SDO mission essentially makes EVE a dedicated flare instrument. 

\subsection{Time-dependent Electron Density Diagnostics}
\label{sec:eve_density}

\begin{figure}[t]
\begin{center}
 \includegraphics[width=0.75\textwidth]{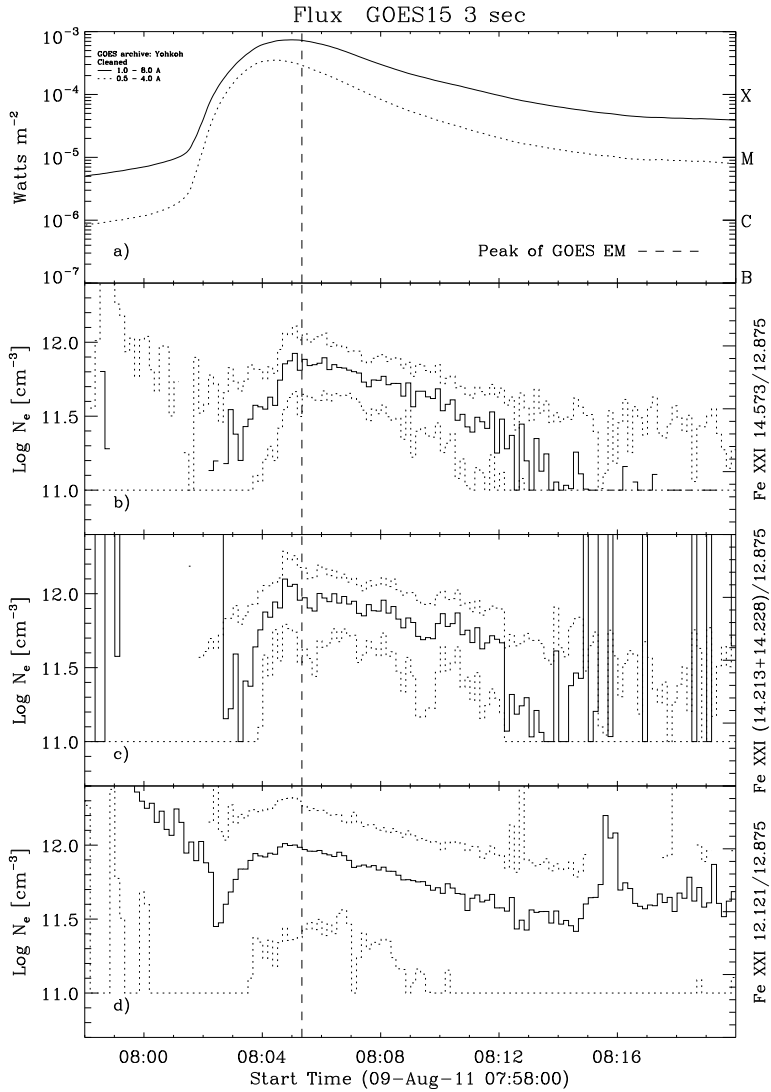} 
 \caption{Top panel: GOES lightcurves of the X6.9 flare on 9 August 2011. Bottom three panels: lightcurves of electron density from three pairs of Fe XXI lines from EVE data. The vertical dashed line marks the peak of the emission measure as determined from GOES data.}
   \label{fig:eve_density}
\end{center}
\end{figure}

EVE offers the ability to obtain values of the coronal electron density by taking the ratio of the flux of two emission lines from the same ionization stage when one of the lines is derived from a metastable transition. \cite{mill12b} identified three pairs of Fe XXI lines within the MEGS-A spectra that provided reliable density measurements for plasma around 12~MK. The line pairs are 121.21\AA/128.75\AA, (142.14\AA+142.28\AA)/128.75\AA, and 145.73\AA/128.75\AA. They found that each of the three pairs gave consistent peak density values of $\sim$10$^{12}$~cm$^{-3}$ (which implied that blending was not an issue) during an X6.9 flare, and that the timing of these peak values agreed well with the time of peak emission measure as determined from broadband GOES measurements (see Figure~\ref{fig:eve_density}). This was repeated for three other X-class flares and consistent values were found, thereby establishing EVE's ability to reliably measure changes in plasma densities in the corona on 10~s timescales during the largest flaring events. 

\subsection{The Anomalous Temporal Behavior of Lyman-alpha Emission During Flares}
\label{sec:eve_lya}

\begin{figure}[t]
\begin{center}
 \includegraphics[width=0.9\textwidth]{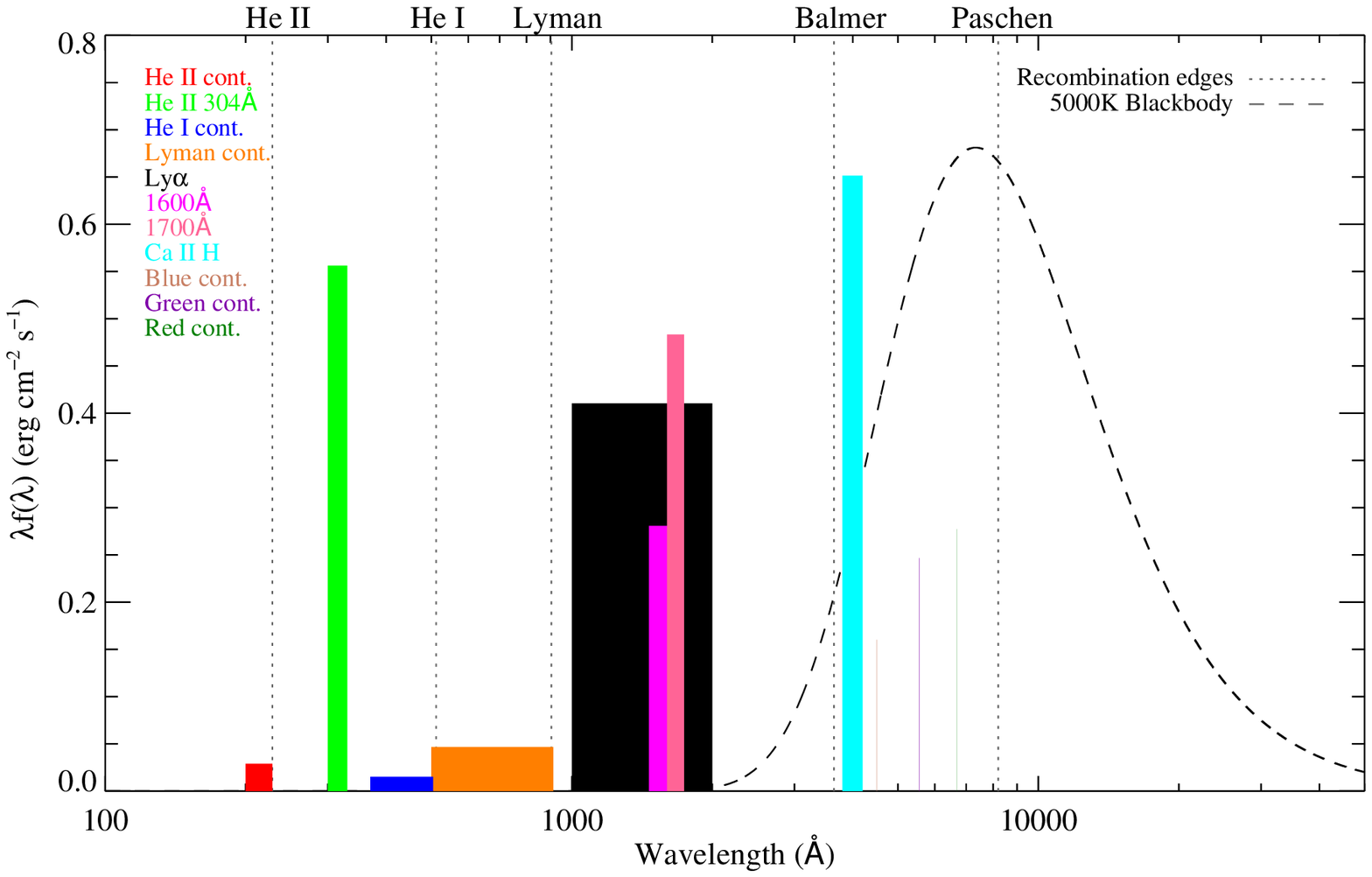} 
 \caption{Spectral Energy Distribution ($\lambda f(\lambda)$) of the flare excess energy. The dashed black curve denotes a blackbody spectrum with a temperature of 5,000~K. The \lya\ line (black histogram) dominated the radiative losses for this event.}
   \label{fig:feb15_spec}
\end{center}
\end{figure}

\begin{figure}[b]
\begin{center}
 \includegraphics[width=0.9\textwidth]{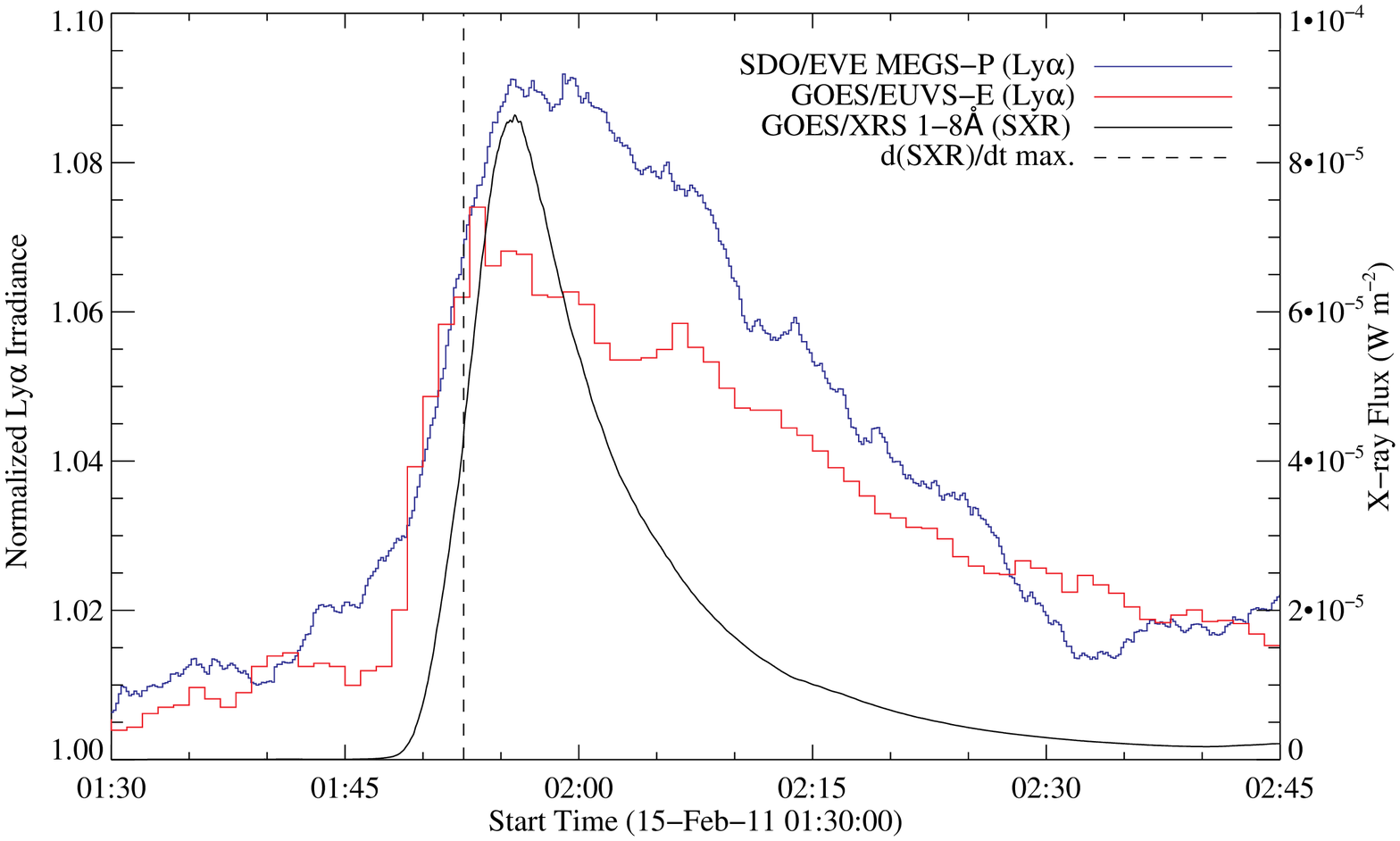} 
 \caption{Plot of normalized \lya\ emission during the 15 February 2011 flare from both EVE MEGS-P (blue) and GOES/EUVS-E (red). Also shown in black is the SXR lightcurve for reference. The \lya\ profile from EVE has a rise time of 10--20 minutes and peaks after the SXR peak, while the \lya\ from EUVS has a rise time of less than 5 minutes and peaks during the impulsive phase. The vertical dashed line denotes the peak time of the SXR derivative.}
   \label{fig:eve_lya}
\end{center}
\end{figure}

The \lya\ transition (2p--1s) of hydrogen is the strongest emission line in the solar spectrum. It is an optically thick line formed in the mid-to-upper chromosphere and recent studies have shown that it to be responsible for radiating away a significant fraction of a flare's energy ($\sim$10\%; \citealt{nusi06,rubi09,mill14}). \lya\ is also known to be a driver of changes in terrestrial ionospheric density in conjunction with SXR photons, although the \lya\ contribution is smaller for flares that occur closer to the solar limb due to absorption along the line of sight \citep{wood06,cham08,qian10}. Understanding variations in \lya\ is therefore a major priority for both solar flare energetics and space weather research. However, despite the importance of \lya\ as a solar diagnostic there are relatively few papers in the literature that discuss changes in \lya\ emission during flares. The few that do come from a variety of different instruments and appear to contradict each other in many aspects (see \citealt{mill15b} for a detailed discussion).

Flare-related enhancements in \lya\ from EVE data were first reported by \cite{mill12a} during the 15 February 2015 X2.2 flare, and a follow-up study revealed \lya\ emission to make up 6-8\% of the total measured radiated losses from the chromosphere for that event (Figure~\ref{fig:feb15_spec}; \citealt{mill14}); as much as all the other measured radiative losses combined. More strikingly though was that the temporal behavior of \lya\ appeared to mimic that of the SXR emission, rather than of the impulsive HXR emission as one might expect for intrinsic chromospheric emission. Figure~\ref{fig:eve_lya} shows the \lya\ time profiles for the 15 February 2011 flare from both EVE (blue) and GOES/EUVS-E (red). The lightcurves of \lya\ from EVE appear to show a gradual, slowly-rising profile, with a rise time of 10--20 minutes. The \lya\ profile from GOES, on the other hand, appears impulsive and bursty with a rise time of just a few minutes. The latter case is what one might expect for a chromospheric plasma heated via Coulomb collisions (e.g., \citealt{brow71}). Furthermore, the MEGS-P \lya\ profiles always appear to peak at, or after, the SXR peak, rather than during the impulsive rise phase, as observed by GOES/EUVS-E.

The reason for the anomalous behaviour of \lya\ observations from EVE is not clear at present. Spectrally and temporally resolved \lya\ profiles from SORCE/SOLSTICE during the 28 October 2003 X17 flare showed that all parts of the line profile peaked during the impulsive phase as expected, ruling out the possibility of opacity effects, or different heating mechanisms affecting different parts of the line, or different layers of the atmosphere responding at different times. Given that the GOES/EUVS-E detector also exhibits  impulsive behaviour during the 15 February 2011 flare even though it has a similar \lya\ filter to EVE, the cause may be due to a data processing algorithm in the EVE pipeline, or the grating that diverts light from the MEGS-B component to MEGS-P may be biasing the observations. In either case, users should exercise caution when interpreting EVE MEGS-P \lya\ data, for either flare science or when injecting \lya\ fluxes into terrestrial atmospheric models. Calculations of \lya\ energetics for the 15 February 2011 flare from GOES/EUVS-E reveal a similar fraction of the total nonthermal electron energy of around 8\%, so it may be more prudent to use these data until the EVE issue is resolved.

\subsection{Deriving The Lyman Continuum Temperature}
\label{sec:lyc_temp_b1}

\begin{figure}[t]
\begin{center}
 \includegraphics[width=\textwidth]{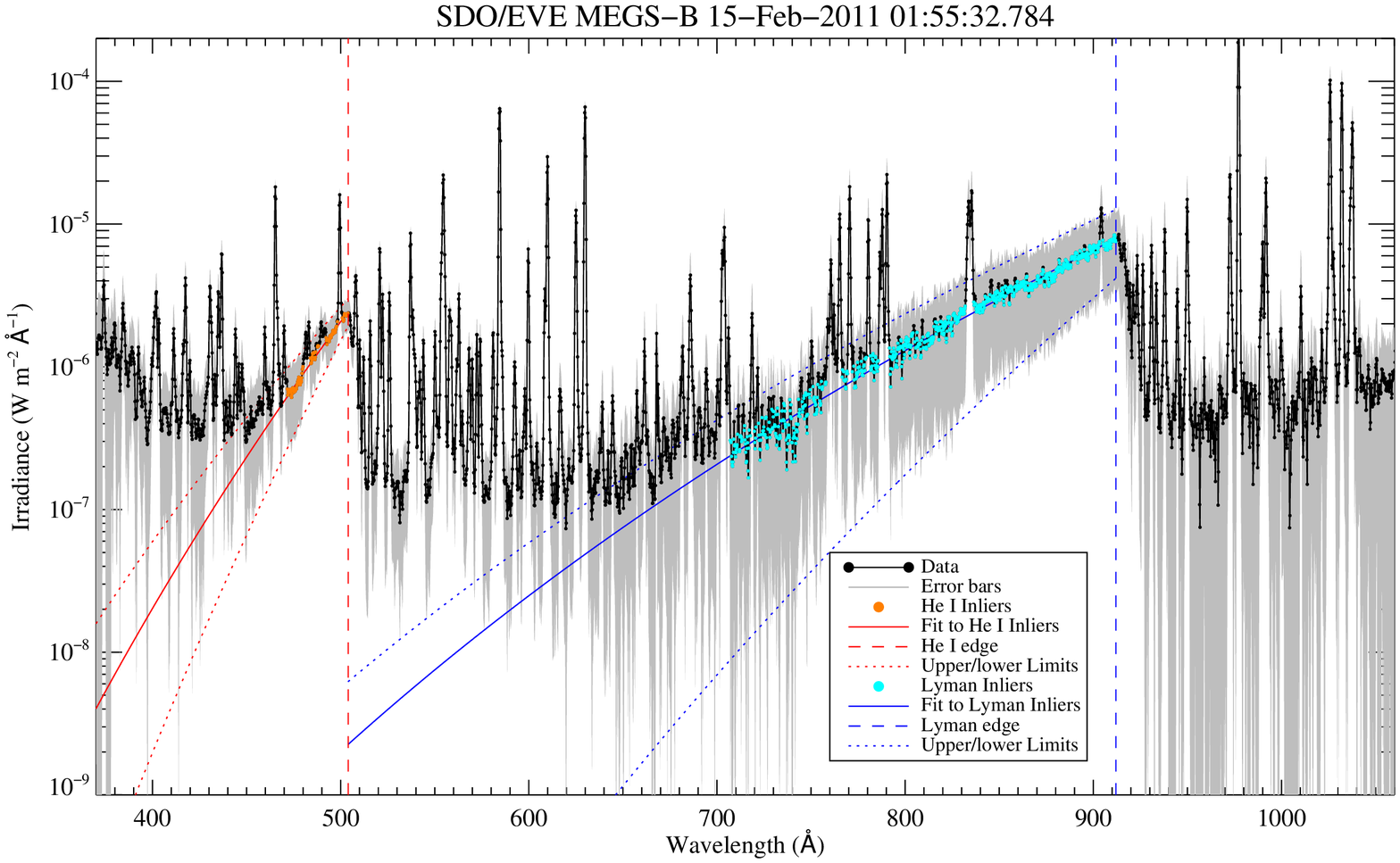}
 \includegraphics[width=\textwidth]{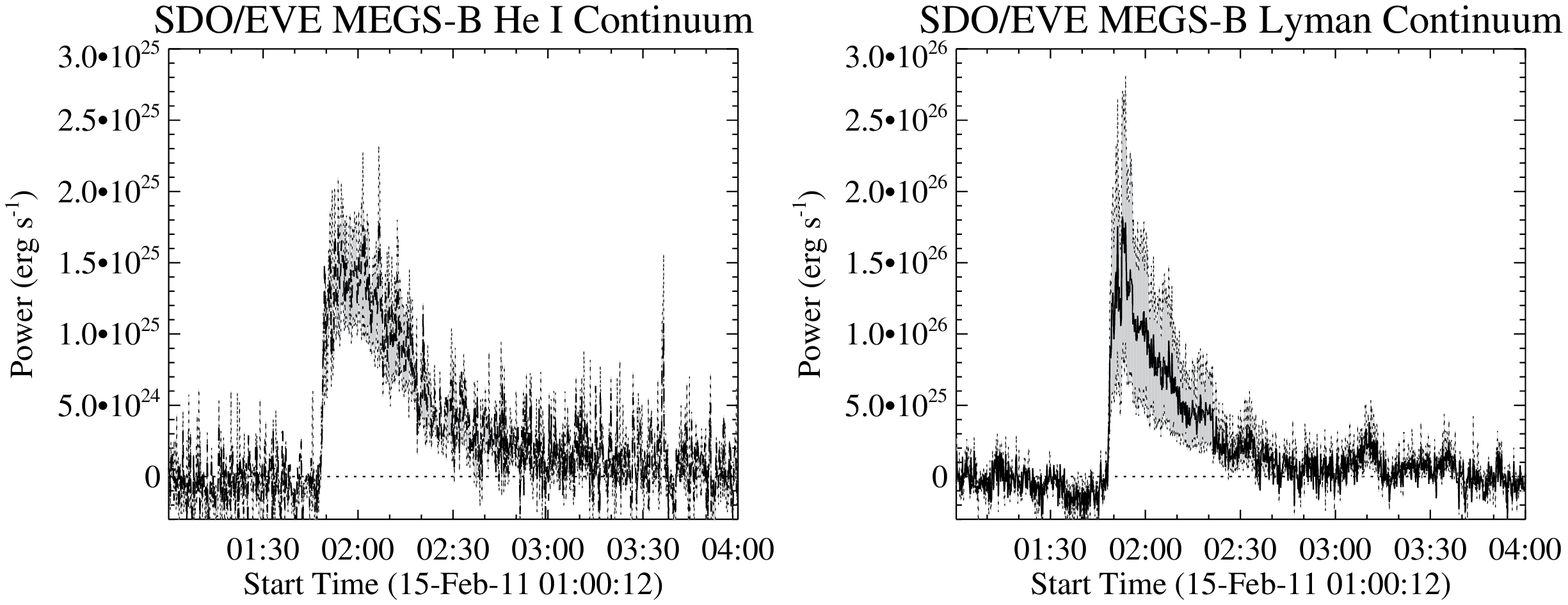}
 \caption{Top: EVE MEGS-B spectra taken near the peak of the X2.2 flare that occurred on 15 February 2011. Overlaid in orange and cyan are the fits to the He~I and Lyman continua, respectively. Bottom: Lightcurves of the He~I (left) and Lyman continua (right) in units of energy for the same event.}
   \label{eve_megsb_cont}
\end{center}
\end{figure}

The MEGS-B component of EVE samples the free-bound (recombination) continua of H~I (LyC) and He~I (top panel of Figure~\ref{eve_megsb_cont}), which are important processes for diagnosing the flaring solar chromosphere. As LyC is formed higher in the atmosphere than the Balmer and Paschen continua it is more sensitive to heating from above. Temporally-resolved measurements of the recombination continua of H and He have now been unambiguously observed during solar flares by EVE \citep{mill12a}. Their time profiles during flares were found to be impulsive (bottom panels of Figure~\ref{eve_megsb_cont}), peaking in concert with the associated HXR emission. This suggested that the increase in continuum emission is due to recombination with free electrons that were liberated during chromospheric heating by nonthermal electrons. MEGS-A also observed the He~II continuum with a recombination edge at 228\AA, but it is inherently weak and only observed during the largest events. It also competes with the underlying free-free continuum, as well as numerous emission lines, adding to the complexity of fitting it and deriving physically meaningful parameters. By integrating under their respective lightcurves, \cite{mill14} found that LyC and the He~I continuum radiated $\sim$1\% and $\sim$0.1\%, respectively, of the total nonthermal electron energy as deduced from RHESSI observations during the 15 February 2011 flare. 

It should therefore be possible to compute the color temperature, $T$, of the continuum and the departure coefficient, $b_1$, of the ground state of H~I as a function of time using the slope of the continuum as derived from the fits to the MEGS-B data. This technique is derived from the Eddington-Barbier relationship and was originally applied to quiet-Sun and active region LyC spectra by \cite{noye70} and \cite{vern72}. \cite{mach78} initially developed this technique and applied it to solar flares, while \cite{ding97} developed it further in anticipation of future LyC observations. They both state that the temperature of the continuum (relative to a blackbody) can be ascertained by taking the slope of the continuum (relative intensity at two distinct wavelengths, $I_{\lambda_1}$ and $I_{\lambda_2}$) using:
\begin{equation}
T = \frac{hc}{k}\left(\frac{1}{\lambda_1}-\frac{1}{\lambda_2}\right)\left[ln\left(\frac{I_{\lambda_2}\lambda_2^5}{I_{\lambda_1}\lambda_1^5}\right)\right]^{-1}
\label{eqn:lyc_temp}
\end{equation}

\noindent
The derived temperature can then be used to yield values of the departure coefficient, $b_1$, given by:
\begin{equation}
\frac{B_{\lambda}(T)}{I_{\lambda}} = \frac{2hc^2}{I_{\lambda}\lambda^5}exp\left(\frac{-hc}{\lambda kT}\right)
\label{eqn:lyc_b1}
\end{equation}

\noindent
where $B_{\lambda}(T)$ is the Planck function.

\cite{mach78} found temperatures in the 8000--9000~K range, with values of $b_1$ lying between 3 and 10. However, these values were only ever obtained once per flare as the spectra were achieved in spectral scan mode. The authors nevertheless concluded that the ground level population of hydrogen is closer to LTE during flares compared to the quiet-Sun ($b_1$=10$^{2}$--10$^{3}$; \citealt{vern72}), and that the height of LyC formation is lower in the solar atmosphere than during quiescent periods. Therefore, despite being energetically very weak, these continua serve as powerful diagnostic tools for determining the depth in the solar atmosphere, and the physical conditions, at which they are formed.

\begin{figure}[t]
\begin{center}
 \includegraphics[width=0.49\textwidth]{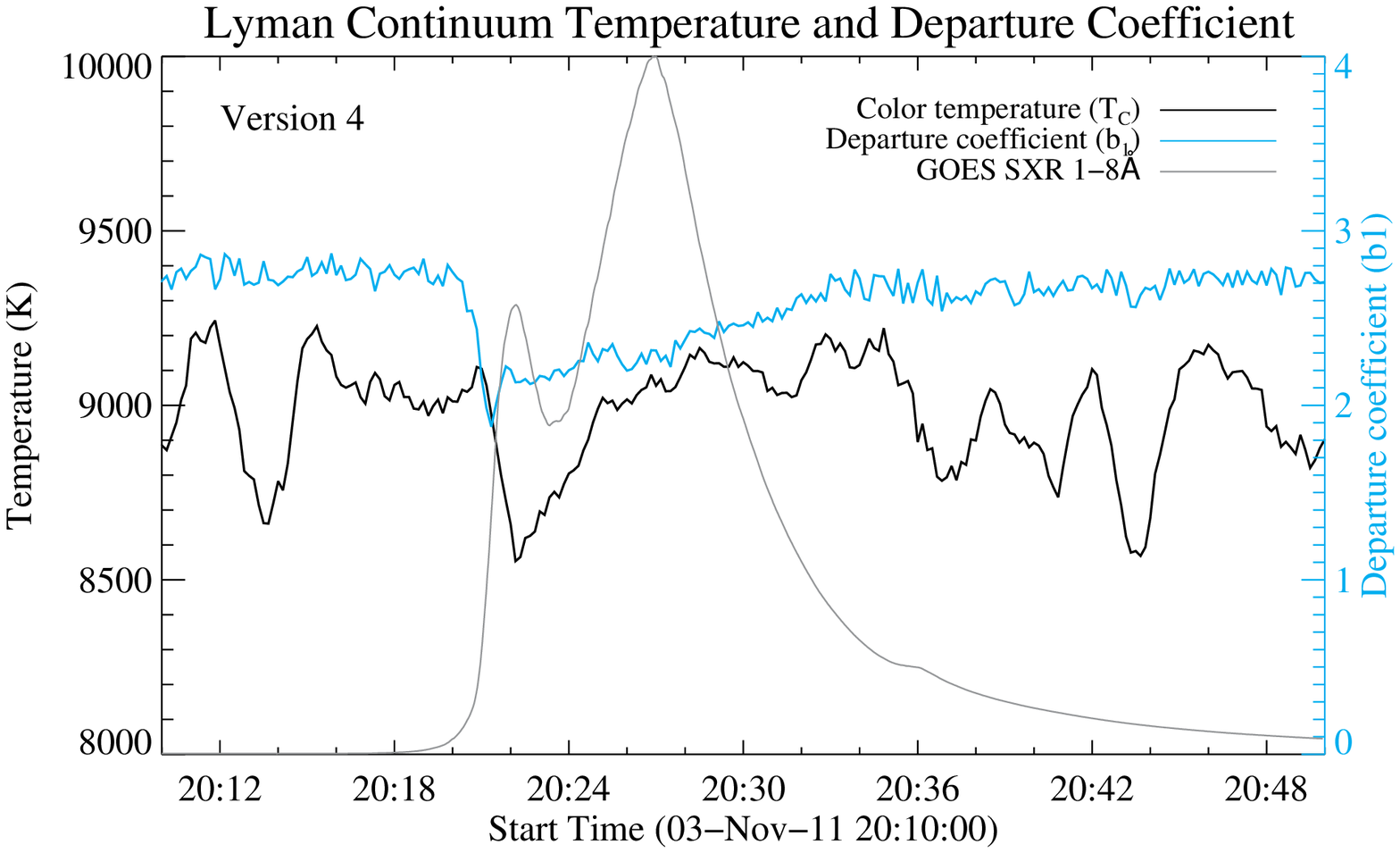} 
 \includegraphics[width=0.49\textwidth]{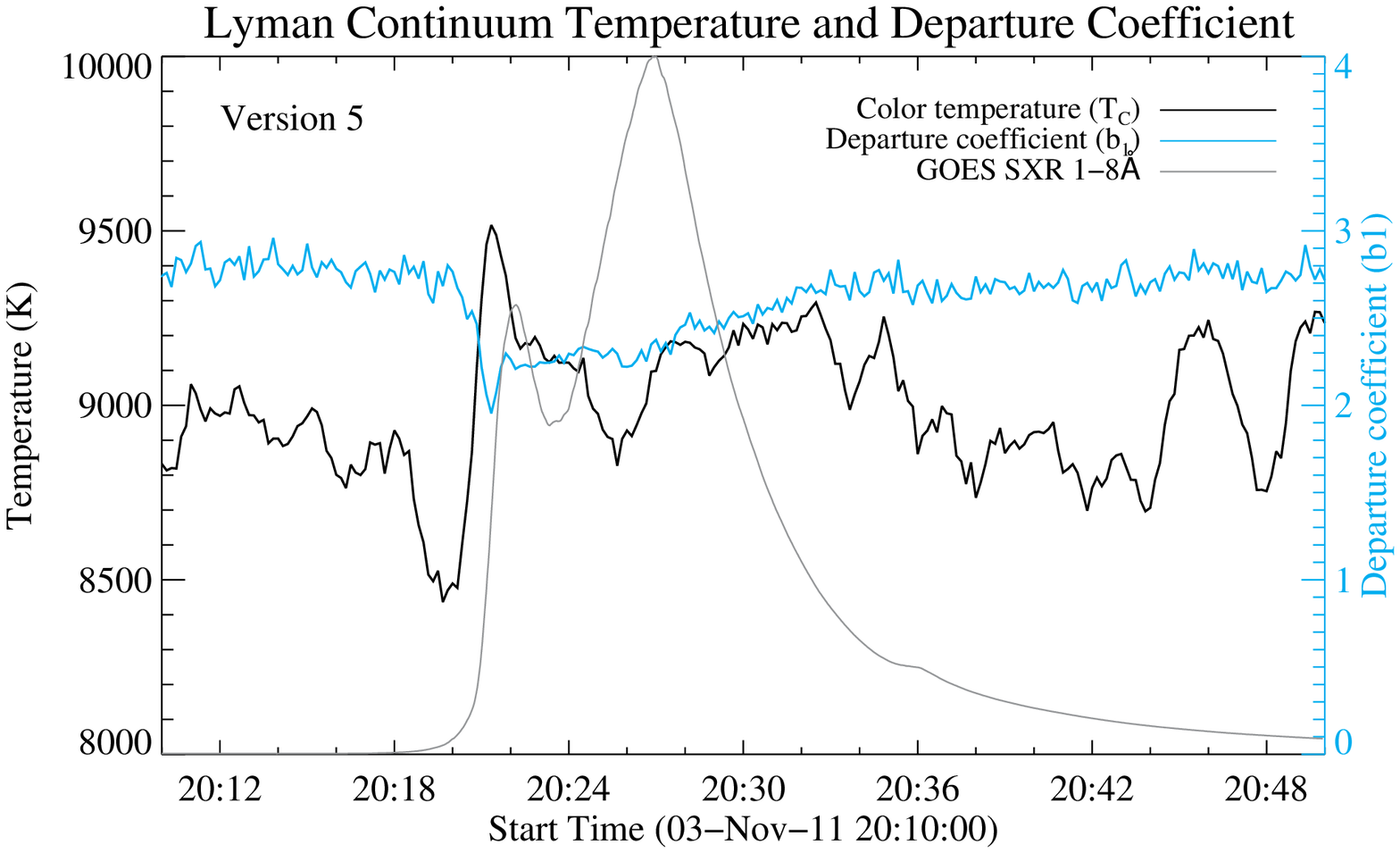} 
 \caption{Plot of the color temperature (black curve) of LyC during the 3 November 2011 flare using versions 4 (left) and 5 (right) of MEGS-B data. The temperature was measured using the Eddington-Barbier relationship by taking the ratio of the fluxes at 800\AA\ and 910\AA\ (smoothed over 60s intervals for clarity). The corresponding departure coefficient, $b_1$, plotted is also plotted (cyan). The GOES SXR lightcurve is overplotted in grey for reference.}
   \label{fig:lyc_temp_b1}
\end{center}
\end{figure}

The left-hand panel of Figure~\ref{fig:lyc_temp_b1} shows a preliminary application of this approach to LyC data from EVE (Version 4) for the 3 November 2011 X1.9 flare. The temperature derived using Equation~\ref{eqn:lyc_temp} (at $\lambda_1$=800\AA\ and $\lambda_2$=910\AA) is plotted in black, although it has been smoothed by a factor of 6 for clarity. This shows a temperature decrease of around 500~K - from 9000~K to 8500~K - in response to the flare, in agreement with \cite{mach78}. Substituting this temperature value back into Equation~\ref{eqn:lyc_b1} allows the computation of the departure coefficient at the head of the LyC ($\lambda=910$\AA). This is overplotted in Figure~\ref{fig:lyc_temp_b1} in cyan, and shows a constant value of $\sim$3 outside of the flare time, but decreases to below 2 around the time of flare onset.

The right-hand panel of Figure~\ref{fig:lyc_temp_b1} shows the same analysis carried out using Version 5 of the EVE data that was released in December 2015. These data now show an {\it increase} in temperature of around 500~K around the time of flare onset in contrast to that obtained using Version 4 data. This discrepancy shows that while the EVE data are stable enough to measure changes in the LyC slope very accurately, the calibration between different versions may not be as reliable. Given that the temperature scales with the inverse fifth power of the wavelength, the slightest changes in calibration can have significant effects on the derived temperature. Perhaps if future versions become more consistent these data can be reexamined. 

\subsection{Lyman Continuum Dimming}
\label{sec:eve_lyc_dim}

\begin{figure}[t]
\begin{center}
 \includegraphics[width=\textwidth]{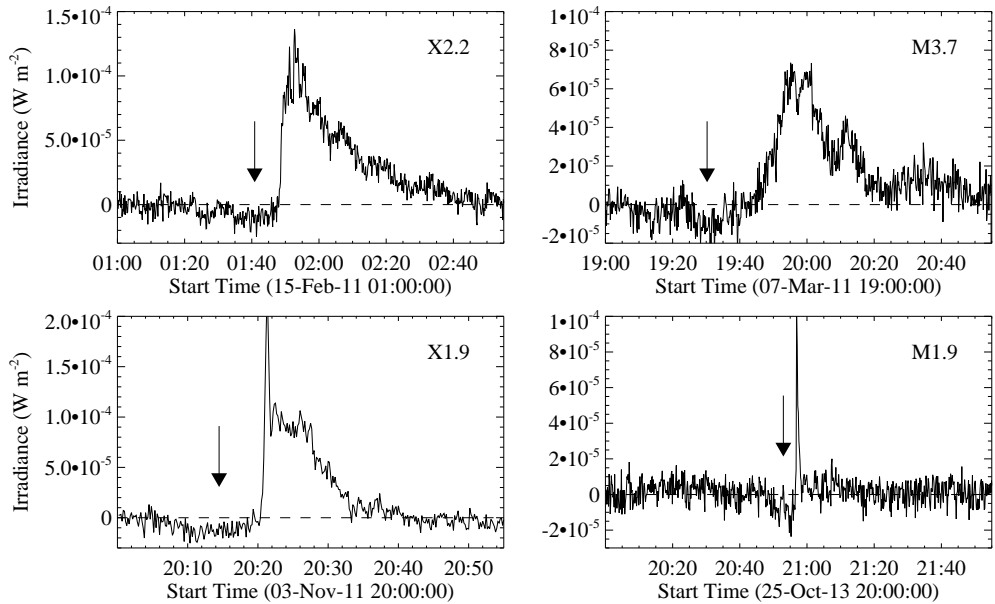} 
 \caption{Background subtracted LyC lightcurve during four major solar flares. The arrows mark the locations of the apparent `dip' prior to the onset of each event, which may be an indicator of continuum dimming due to pre-flare heating and/or increased opacity. Horizontal dotted lines mark the background level.}
   \label{fig:lyc_dim}
\end{center}
\end{figure}

A closer analysis of the 15 February 2011 X-class flare revealed that the background-subtracted LyC lightcurve appeared to drop below the pre-flare level for 10--20 minutes prior to the flare onset. At first glance this would imply that LyC went into absorption; in other words some form of pre-heating or ionization took place before the main energy release. This is analogous to the so-called ``black light flares'' (see \citealt{vand94} and references therein). A small number of other events have since shown a similar behavior (see Figure~\ref{fig:lyc_dim}). This ``continuum dimming'' has been predicted for the Balmer and Paschen continua in numerical simulations by \cite{abbe99} and \cite{allr05}, although it only lasted for $<$0.1~s. To date, the only apparent evidence that supports these predictions has been from \cite{giam82} who observed Paschen dimming prior to a stellar flare on EQ Pegasi that lasted for 2.7 minutes. \cite{abbe99} explained that this dimming is due to an increase in the population density of higher order bound states of hydrogen. Balmer and Paschen photons generated in the photosphere therefore get absorbed in the chromosphere resulting in an initial decrease in emission. However, \cite{abbe99} also state that because of the reduced populations of the ground level of hydrogen, no such absorption should be apparent for LyC. Clearly, the timescales of the dimming vary greatly between observations and theory, but bear in mind that the models are generated for a single, one-dimensional flux tube whereas the observations are from an extensive three-dimensional flare arcade.

\subsection{SDO/AIA Continuum Contributions}
\label{sec:eve_aia}

\begin{figure}[h]
\begin{center}
 \includegraphics[width=0.9\textwidth]{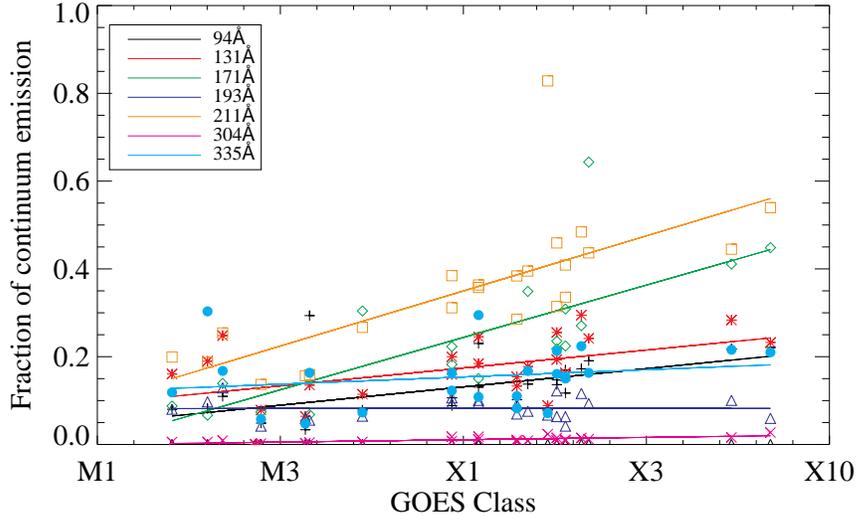} 
 \caption{Plots of the continuum contribution to each of the AIA channels as a function of GOES class at the time of peak SXR emission in 24 events. The tick marks on the $x$-axis denote steps of 0.1 dex in the 1--8\AA\ flux. From \cite{mill13}.}
   \label{fig:eve_aia}
\end{center}
\end{figure}

The Atmospheric Imaging Assembly (AIA) instrument on SDO has been taking full-disk, high-resolution EUV images of the solar atmosphere at high cadence for almost 6 years. Each passband is broadly assumed to focus on a single, coronal emission line formed between 1--7 MK. During solar flares some channels may also include emission from lines formed at even higher temperatures ($>$10 MK). However, using EVE data, \cite{mill12a} showed that the free-free continuum, which spans the wavelength range that contains the AIA channels, also becomes significantly enhanced during flares. A follow-up study by \cite{mill13} aimed to quantify how much this continuum emission contributed to each of the AIA passbands during flares by convolving both the EVE spectra and fits of the free-free continuum with the AIA response functions for a number of flares of varying magnitudes. While most channels were found to contain less than 20\% free-free emission, the 171\AA\ and 211\AA\ channels were found to contain up to 50\% free-free emission during the largest events (see Figure~\ref{fig:eve_aia}). This is because these two channels do not contain high temperature line emission and therefore continuum emission makes up a larger portion of the observed emission as flare temperature surpass 10~MK. Knowledge of which emission processes dominate the AIA images under different conditions are crucial for interpreting the data correctly.

\section{Conclusions}

Despite being designed to monitor changes in the solar EUV flux at Earth over a range of timescales (seconds and minutes during flares; days and weeks as active regions appear and disappear over the course of a solar rotation; and months and years over the course of a solar cycle), the EUV Variability Experiment also offers flare scientists a wealth of diagnostic information with which to probe the plasma conditions in the flaring solar atmosphere. This means that EVE is not only capable of quantifying changes in the geoeffective solar irradiance, but its data can also be used to help understand the physical mechanisms responsible for driving these changes. The findings outlined in this paper that have been obtained from EVE data to date is by no means exhaustive. Many other authors have capitalised on the uniqueness of EVE data with the promise of more revelations on flare characteristics in the future. Despite the loss of MEGS-A (and SAM) in May 2014, MEGS-B and -P continue to operate at a reduced duty cycle and have observed more than 100 flares of GOES class M1 or greater since launch, often in conjunction with other missions such as RHESSI, Hinode and IRIS. The emission lines and continua that are observed by EVE - even those that are optically thick - are directly reproducible by radiative hydrodynamic codes such as RADYN \citep{allr05,allr15}. This allows us to explore how changes in the distribution of nonthermal electrons, the introduction of thermal conduction fronts, return currents and even proton beams, affect the response of the solar atmosphere to an injection of energy, and how this ultimately causes variations in the Sun's EUV output. Coordinating EVE observations with other space- and ground-based instruments can help address long-standing issues on energy release and transport during solar flares, and should be strongly encouraged.

\end{document}